\documentstyle[preprint,prb,aps]{revtex}
\begin{document}
\draft
\noindent
\begin{centering}

{\Large CONNECTION BETWEEN THE SLAVE-PARTICLES AND X-OPERATORS PATH-INTEGRAL 
REPRESENTATIONS. A NEW PERTURBATIVE APPROACH.\\}

\vspace{1.cm}
{\bf A.Foussats$^{*}$, A. Greco$^{*}$, C.Repetto$^{*}$, 
O.P.Zandron$^{*}$ and O.S.Zandron$^{*}$}\\

\vspace{0.3cm}

{\footnotesize $*$ Miembros del Consejo Nacional de Investigaciones 
Cient\'{\i}ficas y T\'ecnicas - Argentina.}\\
\vspace{1.cm}
{\em Facultad de Ciencias Exactas Ingenier\'{\i}a y Agrimensura de la UNR,\\
Av.Pellegrini 250 - 2000 Rosario - Argentina.}\\
\end{centering}

\begin{abstract}

\begin{centering}
{\bf Abstract\\}
\end{centering}

In the present work it is shown that the family of first-order
Lagrangians for the t-J model and the corresponding correlation
generating functional previously found can be exactly mapped into the
slave-fermion decoupled representation. Next, by means of the
Faddeev-Jackiw symplectic method, a different family
of Lagrangians is constructed and it is shown how the
corresponding correlation generating functional can be mapped
into the slave-boson representation. Finally, in order to define
the propagation of fermion modes we discuss two
alternative ways to treat the fermionic sector in the
path-integral formalism for the t-J model.

\end{abstract}


\narrowtext
 
\newpage 

\section{Introduction}

Accounting the relevance in describing the behavior of strongly 
correlated electron systems, a renovate interest 
in the study of the supersymmetric generalisations of the Hubbard 
model appears in the last years. A complete review on strongly correlated 
electron systems together with its connection to high-$T_{c}$ is given 
in Ref.[1].

The Hubbard models based on the superalgebras spl(2,1), osp(2,2) or su(2,2)
have been formulated by using several
approaches\cite{2,3,4,5,6}. For instance, as it was
suggested\cite{7,8,9} the superalgebra spl(2,1) could be useful 
to study the model in the limit of infinite on-site
repulsion and with infinite-range hopping between all sites\cite{10}.  

Many problems about correlated electron systems were treated in the 
framework of the decoupled slave-particle representations. Two
of them are the most important: 
the slave-boson and the slave-fermion representations. The first
one privileges the fermion dynamics, and therefore the slave-boson
representation seems to be more ade\-qua\-ted to describe a Fermi
liquid state\cite{11,12}. Instead, the slave-fermion representation
seems to give a good response when the system is closed to the 
antiferromagnetic order\cite{13,14}. 

An important question in order to understand the physics of the high-$T_{c}$ 
superconductors, is to solve how to go from one representation to the other. 
This is because in high-$T_{c}$ superconductors, both 
the Fermi liquid and the magnetic order states seem to be present.

On the other hand, one of the main problems appearing in these kind of 
models is to define the
dynamics of fermions in the constrained Hilbert space, when the
double occupancy of lattice sites is excluded. In this case also a
convenient representation is given in terms of slave-particles\cite{15}.

As it is known the slave-particle models exhibit a local gauge invariance 
which is destroyed in the mean field approximation.
This local gauge invariance has associated a first-class
constraint which is difficult to handle in the path-integral formalism. 
 
Another possibility to attack the problem was given in Refs.[16,17] 
by using gene\-ra\-lised coherent states in the framework of the functional 
integral formalism.
 
Recently, the t-J model was analysed in the context of the
path-integral formalism\cite{18,19}. Our starting point was the
construction of a particular family of first-order cons\-trai\-ned Lagrangians
by using the Faddeev-Jackiw (FJ) symplectic method\cite{20}, in
the supersymmetric version\cite{21,22}. 
In this approach any decoupling is used, but the field variables
are directly the Hubbard $X$-operators which verify the
superalgebra spl(2,1). In this way, we always work with the real physical
excitations. 

Next, by using path-integral techniques, the correlation
generating functional and the effective Lagrangian were constructed.
Moreover, we have proved that our path-integral representation can be
directly related with that found in Ref.[23].

As mentioned above, one of the interesting and non completly
solved problem present in this constrained system, is to study 
the fermionic sector when the double occupancy of lattice sites
is excluded. In particular the role of the fermionic cons\-traints,
and so the fermionic dynamics in the constrained Hilbert space
is a crucial problem that must be investigated. 

Therefore, one of the main purposes of the present paper is to go deep into the
discussion of the different alternatives which allows us to define the 
fermionic propagator in the t-J model.
 
In Ref.[19], in the framework of the perturbative formalism the
Feynman rules with appropriate propagators and vertices were found.
In particular, a discussion on the fermionic propagator was also given.

In order to continue with the study of the fermionic propagation, different 
alternatives are exposed in the present paper. We show that  
there is another way that the one given in Ref.[19] to obtain 
fermionic propagation. This is done by working inside 
the path-integral and integrating out the two delta functions 
on the fermionic constraints.

Moreover, other interesting point is to check our formalism with
those obtained by means of the slave-particle representations. More
precisely, we will show how our path-integral expression for the
partition function (see equation (4.1) of Ref.[18]), written in terms of the
Hubbard operators, can be mapped in the partition function
coming from the slave-fermion representation. 

On the other hand, by following the FJ symplectic method, 
it is possible to show that a new family of first-order constrained 
Lagrangians written in terms of the Hubbard 
$X$-operators exists. This family of classical Lagrangians is able to 
reproduce the Hubbard $X$-operators commutation rules verifying the graded 
algebra spl(2,1). As it can be shown this family of Lagrangians, totally
cons\-trai\-ned in the boson-like Hubbard $X$-operators, can be mapped into 
the slave-boson representation.
 
The paper is organized as follow. In section II, the main
results of Refs.[18,19] are collected.
In section III, by analysing the change in the constraints
structure of the t-J model, it
is showed how starting from our path-integral expression 
for the partition function previuosly found, the partition function coming 
from the decoupled slave-fermion representation can be recovered.  
In section IV, by using the FJ symplectic method, a different
family of first-order constrained Lagrangians is found. This
family of Lagrangians corresponds to the situation in which the bosons are
totally constrained. In such conditions it is possible
to show how the correspondent partition function can be mapped in the 
partition function coming from the slave-boson representation. In section V, 
two alternative ways to define the fermion propagation are studied. 
In section VI, conclusions are given.

\section{Preliminary and Definitions}

In the t-J model the three possible states on a lattice site are 
$\mid\alpha>$ = $\mid 0>$, $\mid +>$, $\mid ->$. These states correspond
respectively to an empty site, an occupied site with an electron of spin-up, 
or an occupied site with an electron of spin-down. Double occupancy is 
forbidden in the t-J model. In terms of these states the Hubbard 
${\hat X}$-operators are defined as

\begin{equation}
{\hat{X}}^{\alpha \beta}_{i} = \mid i \alpha>< i \beta\mid\; .
\eqnum{2.1}  
\end{equation}

In Eq.(2.1), when one of the index is zero and the other different 
from zero, the corresponding ${\hat X}$-operator is fermion-like, otherwise 
boson-like. 

The Hubbard ${\hat X}$-operators verify the following graded commutation 
relations

\begin{equation}
[{\hat{X}}_{i}^{\alpha \beta}\,,\,{\hat{X}}_{j}^{\gamma \delta}]_{\pm}
=\delta_{ij}(\delta^{\beta \gamma}{\hat{X}}_{i}^{\alpha \delta}\; {\pm}\;
\delta^{\alpha \delta}{\hat{X}}_{i}^{\gamma \beta})\; , 
\eqnum{2.2}
\end{equation}

\noindent
where the $+$ sign must be used when both operators are
fermion-like, otherwise it corresponds the $-$ sign, and $i$, $j$ denotes 
the site indices. 

We assume that the family of classical constrained first-order Lagrangians 
in terms of the Hubbard ${\hat X}$-operators can be written as follows 

\begin{equation}
L = a_{\alpha \beta}(X){\dot{X}}^{\alpha \beta} - {\bf V^{(0)}}\;.
\eqnum{2.3}
\end{equation} 

In the FJ language\cite{20} the symplectic potential ${\bf V^{(0)}}$ 
is defined by

\begin{equation}
{\bf V^{(0)}} = H(X) + \lambda^{a}\Omega_{a}\;, 
\eqnum{2.4}
\end{equation}

\noindent
where $\lambda^{a}$ are appropriate Lagrange multipliers, and so 
the constraints $\Omega_{a}$ are defined by

\begin{equation}
\Omega_{a} = \frac{\partial{\bf V}^{(0)}}{\partial{\lambda^{a}}}\;.
\eqnum{2.5}
\end{equation}

Therefore, the symplectic supermatrix associated to the Lagrangian (2.3) 
can be formally written\cite{22}

\begin{eqnarray}
M_{AB} =
\left( \matrix {
\frac{\partial a_{\gamma \delta}}{\partial X^{\alpha \beta}}
- (-1)^{\mid \alpha \beta \mid \mid \gamma \delta\mid}
\frac{\partial a_{\alpha \beta}}{\partial X^{\gamma \delta}}
&\frac{\partial\Omega_{b}}{\partial X^{\alpha \beta}}   \cr
- (-1)^{\mid a \mid \mid \gamma\delta\mid}
\frac{\partial\Omega_{a}}{\partial X^{\gamma \delta}}    &0    \cr }
\right) \; , \eqnum{2.6}
\end{eqnarray}

\noindent
where the compound indices $A = \{{(\alpha \beta)}, a \}$ and 
$B =\{{(\gamma \delta)}, b\}$ run in the different ranges of the complete
set of variables defining the extended configuration space, and 
$\mid A \mid$ indicates the Fermi grading.

Following Ref.[18], our starting point is to consider the following 
partition function for the t-J model written in terms of the four boson-like 
operators $(X^{+-}\;,\;X^{-+}\;,\;X^{++}\;,\;X^{--})$  and the four 
fermion-like operators $(X^{0+}\;,\;X^{0-}\;,\;X^{+0}\;,\;X^{-0})$

\begin{equation}
Z = \int {\cal D}X^{\alpha \beta}_{i}\; \delta(\Omega_{i1})\;
\delta(\Omega_{i2})\; \delta(\Xi_{i3})\;\delta(\Xi_{i4})\; 
(sdet M_{AB})_{i}^{\frac{1}{2}}\;exp\;i \int dt\; L(X,\dot{X}) \;,
\eqnum{2.7}
\end{equation}

\noindent
where $L(X,\dot{X})$ is given by

\begin{eqnarray}
L(X,\dot{X}) & = & i\sum_{i}\frac{(1 + \rho_{i})u_{i} - 1}{(2 - v_{i})^{2} 
- 4\rho_{i} - u^{2}_{i}}
\left(X^{- +}_{i} {\dot X}^{+ -}_{i} - X^{+ -}_{i} {\dot X}^{- +}_{i}\right)
\nonumber \\ 
& + & \frac{i}{2} \sum_{i, \sigma} \left({\dot X}_{i}^{0 
\sigma}X_{i}^{\sigma 0} + {\dot X}_{i}^{\sigma 0} X_{i}^{0 \sigma}\right)
- \mu \sum_{i, \sigma}{X^{0 \sigma}_{i}}{X^{\sigma 0}_{i}} - H_{t-J}(X)\;.
\eqnum{2.8}
\end{eqnarray}

\noindent
where $u_{i} = X_{i}^{+ +} - X_{i}^{- -}$ and 
$v_{i} = X_{i}^{+ +} + X_{i}^{- -}$.

The Greek indices ${\alpha, \beta}$ takes the values $\{+ , - , 0\}$, 
the index ${\sigma}$ takes the values $\{+ , - \}$, and $H_{t-J}(X)$ 
is the usual t-J Hamiltonian 

\begin{equation}
H_{t-J} =  \sum_{i,j,\sigma} \; t_{ij}\; X^{\sigma 0}_{i}
X^{0 \sigma }_{j} + \frac{1}{4} \sum_{i,j, \sigma, \bar{\sigma}}
J_{ij}\; X^{\sigma {\bar{\sigma}}}_{i}
X^{\bar{\sigma} \sigma}_{j} - \frac{1}{4} \sum_{i,j, \sigma, \bar{\sigma}}
J_{ij}\; X^{\sigma \sigma}_{i}
X^{\bar{\sigma}{\bar{\sigma}}}_{j}\;, 
\eqnum{2.9}
\end{equation}

\noindent
besides, in equation (2.8) a term depending on the chemical potential $\mu$ 
was added.

In Eq.(2.7) $sdet M_{AB}$ is the superdeterminant of the
symplectic supermatrix $M_{AB}$ defined in (2.6), and the bosonic and 
fermionic constraints at each site $i$ are given respectively by

\begin{equation}
\Omega_{i 1} = X^{++}_{i} + X^{--}_{i} + \rho_{i} - 1 = 0\;, 
\eqnum{2.10a}
\end{equation}
 
\begin{equation}
\Omega_{i 2} = X^{+ -}_{i} X^{- +}_{i} + \frac{1}{4}
(X^{++}_{i} - X^{--}_{i})^{2} - [1 - \frac{1}{2}(X^{++}_{i} + X^{--}_{i})]^{2}
+ \rho_{i} = 0\;, 
\eqnum{2.10b}
\end{equation}

\begin{equation}
\Xi_{i 3} =  X^{0+}_{i} X^{+-}_{i} - X^{0-}_{i} X^{++}_{i} = 0\;, 
\eqnum{2.10c}
\end{equation}

\begin{equation}
\Xi_{i 4} =  X^{+0}_{i} X^{-+}_{i} - X^{-0}_{i} X^{++}_{i} = 0\;. 
\eqnum{2.10d}
\end{equation}

\noindent
where was defined $\rho_{i} = X^{0+}_{i} X^{+0}_{i} + X^{0-}_{i} X^{-0}_{i}$. 

In equation (2.8), the Lagrangian 
coefficients as well as the constraints (2.10) were determined by using the 
FJ symplectic method with the condition
to reproduce at the classical level the generalised FJ brackets or graded 
Dirac brakets of the t-J model (see Ref.[18]). 

In particular the constraint (2.10a) deduced by consistency, is the 
completeness condition which must be verified by the Hubbard
${\hat X}$-operators and plays an important physical role in this
constrained model as it will be seen later on. At this stage it is important 
to remark that in the Dirac language\cite{24} the two
bosonic constraints (2.10a,b) are second-class one. 

Now, it is useful to write the boson-like Hubbard $X$-operators
in terms of the real components $S_{\alpha}$ ($\alpha
= 1,2,3$) of a vector field ${\bf S}$ and the fermion-like Hubbard 
$X$-operators in terms of suitable component spinors 
(Grassmann variables)\cite{18,23} 

\begin{equation}
X_{i}^{+ +} = \frac{1}{2 s} (1 - \rho_{i})(s + S_{i 3}) \;,
\eqnum{2.11a} 
\end{equation}

\begin{equation}
X_{i}^{- -} = \frac{1}{2 s} (1 - \rho_{i})(s - S_{i 3}) \;,
\eqnum{2.11b} 
\end{equation}

\begin{equation}
X_{i}^{+ -} = \frac{1}{2 s} (1 - \rho_{i})(S_{i 1} + iS_{i 2}) \;,
\eqnum{2.11c} 
\end{equation}

\begin{equation}
X_{i}^{- +} = \frac{1}{2 s} (1 - \rho_{i})(S_{i 1} - iS_{i 2}) \;,
\eqnum{2.11d} 
\end{equation}

\begin{equation}
X_{i}^{- 0} =  \Psi_{i +}\;,  \hspace{2cm} X_{i}^{0 -} =
\Psi_{i +}^{*}\;,
\eqnum{2.11e} 
\end{equation}

\begin{equation}
X_{i}^{+ 0} = \Psi_{i -}\;,  \hspace{2cm} X_{i}^{0 +} = \Psi_{i -}^{*}\;,
\eqnum{2.11f} 
\end{equation}
 
\noindent
where $s$ is a constant and the hole density in the new
variables writes $\rho_{i} = 
\Psi_{i +}^{*} \Psi_{i +} + \Psi_{i -}^{*} \Psi_{i -}$. 
Accounting the fermionic constraints (2.10) it results 
$(1 - \rho_{i})(1 + \rho_{i}) = 1$.

The real vector field ${\bf S}$ can be identified with the spin 
only when $\rho = 0$, i.e., in the pure bosonic case. 

By using the second-class constraints (2.10a) and after the change of 
variables is made, the partition function takes the form 

\begin{equation}
Z  =  \int  {\cal D} S_{i1} \; {\cal D} S_{i2}\; {\cal D} S_{i3}\; 
{\cal D}{\Psi}_{i \sigma}\; 
{\cal D}{\Psi}^{*}_{i \sigma}\; 
{\cal D}{\lambda_{i}}\; {\cal D}{\xi_{i}}\; {\cal D}{\xi_{i}^{*}}\;
(sdet M_{AB})_{i}^{\frac{1}{2}}\;
\left(\frac{\partial X}{\partial S}\right)_{i}\;
exp\;(i \int\; dt \; L_{eff})\;.
\eqnum{2.12}
\end{equation}

\noindent 
where the quantity $\left(\frac{\partial X}{\partial S}\right)_{i}$ is 
the super Jacobian of the transformation (2.11).

The effective Lagrangian $L_{eff}$ defined in Eq. (2.12), in terms of the new
variables reads  
 
\begin{eqnarray} 
L_{eff} & = & \frac{1}{2s} \sum_{i} \frac{S_{i1}{\dot S}_{i2} 
- S_{i2}{\dot S}_{i1}}{s + S_{i3}} + i \sum_{i, \sigma} 
{\dot{\Psi}^{*}}_{i \sigma} {\Psi_{i \sigma}} +
\mu \sum_{i, \sigma}{\Psi_{i \sigma}}{\Psi^{*}_{i \sigma}} - H_{t-J}
\nonumber \\
& + & \sum_{i}\left[\lambda_{i}(S_{i1}^{2} + S_{i2}^{2} + S_{i3}^{2} 
- s^{2})
+ \xi^{*}_{i} \left(\Psi_{-}(S_{i1} - iS_{i2}) - \Psi_{+}(s + S_{i3})
\right) \right. \nonumber \\ 
& + & \left. \left(\Psi_{-}^{*}(S_{i1} + iS_{i2}) - \Psi_{+}^{*}(s + S_{i3})
\right) \xi_{i} \right] \;,
\eqnum{2.13}
\end{eqnarray}

\noindent
where the Hamiltonian $H_{t-J}$ of the t-J model writes

\begin{equation}
H_{t-J} =  \sum_{i,j,\sigma} t_{ij} \Psi_{i \sigma}\Psi_{j \sigma}^{*}
+\frac{1}{8 s^{2}}\sum_{i,j} J_{ij}(1 - \rho_{i})(1 - \rho_{j}) \left[
S_{i1}S_{j1} + S_{i2}S_{j2} + S_{i3}S_{j3} - s^{2}\right]\;.  
\eqnum{2.14}
\end{equation}

In equation (2.13) the parameters $\lambda_{i}$, $\xi^{*}_{i}$
and $\xi_{i}$ are respectively suitable bosonic and fermionic Lagrange 
multipliers. 

At this stage it is important to remark that our Lagrangian
formalism is independent of the underlying lattice dimension.

In the next section we are going to analyse the equation (2.7)
in the framework of the decoupled slave-particle representations.

\section{Slave-particle representations.}

The standard way to construct the classical Hamiltonian formulation for
slave-particle models and subsequently give the canonical quantisation is
developed in Ref.[15] . The starting   
point is to consider the general classical first-order Lagrangian for $n$
bosonic fields $b_{a}$ and $m$ fermionic fields $f_{b}$
defined on a lattice  

\begin{equation}
L(b^{\dag}_{a},b_{a}, f^{\dag}_{b},f_{b}) =
\frac{i}{2}\sum_{i , a}(b^{\dag}_{i a} \dot{b}_{i a} -
\dot{b^{\dag}}_{i a} b_{i a}) +
\frac{i}{2}\sum_{i , b}(f^{\dag}_{i b}\dot{f}_{i b} -
\dot{f}^{\dag}_{i b} f_{i b}) - H(b^{\dag}_{a},
b_{a}, f^{\dag}_{b},f_{b})\;.
\eqnum{3.1}
\end{equation}

Both, bosons and fermions fields, are submitted to the slave-particle
first-class constraint at each lattice site $i$ 

\begin{equation}
\Omega_{i} = \sum_{a} b^{\dag}_{i a}b_{i a} 
+ \sum_{b} f^{\dag}_{i b} f_{i b} - 1 = 0\;.
\eqnum{3.2}
\end{equation}

Looking at the equations (3.1) and (3.2) it can be seen that when the index 
$a$ takes the values $\pm$ and the index $b$ takes only one value, the
six fields (four boson and two fermion fields) define the slave-fermion 
representation. On the contrary, when the index $a$ takes
only one value and the index $b$ takes the values $\pm$, the
six fields (two boson and four fermion fields) define the 
slave-boson representation.

With the aim to confront our results with others previously obtained, in 
this section we consider the 
slave-fermion representation. In particular it is possible to confront the
correlation generating functional (2.7) with that obtained
from the slave-fermion representation, and this relation is not 
trivial. 

In our approach all the constraints are in the
Dirac picture\cite{24} second-class, while in the slave-particle
representations the constraint (3.2) is first-class. Thus, when the
Hubbard $X$-operators are decoupled a local gauge symmetry is
made evident.

The starting point is the correlation generating functional
(2.7) with the Lagrangian (2.8).

By computing the $(sdet M_{AB})_{i}^{\frac{1}{2}}$
appearing in equation (2.7) we find

\begin{equation}
(sdet M_{AB})_{i}^{\frac{1}{2}}  = - i \frac{(1 + \rho_{i})}{X_{i}^{+ +}}\;. 
\eqnum{3.3}
\end{equation}

\noindent
where $\rho_{i}$ evaluated on the constraints is written $\rho_{i} = 
\frac{X_{i}^{0 +} X_{i}^{+ 0}}{X_{i}^{+ +}}$.

Integrating out the fields components $X_{i}^{- -}$, $X_{i}^{0 -}$ 
and $X_{i}^{- 0}$
by using the delta functions on the constraints written as follows

\begin{equation}
\delta(\Omega_{i 2}) = \delta(X_{i}^{+ +} X_{i}^{- -} - X_{i}^{+ -} 
X_{i}^{- +}) 
= \frac{1}{X_{i}^{+ +}}\delta(X_{i}^{- -} - \frac{X_{i}^{+ -} 
X_{i}^{- +}}{X_{i}^{+ +}})\;,  
\eqnum{3.4a}
\end{equation}

\begin{equation}
\delta(\Xi_{i 3}) = \delta(X_{i}^{0 +} X_{i}^{+ -} - X_{i}^{0 -} X_{i}^{+ +}) 
= X_{i}^{+ +}\delta(X_{i}^{0 -} - \frac{X_{i}^{0 +} X_{i}^{+ -}}
{X_{i}^{+ +}})\;,  
\eqnum{3.4b}
\end{equation}

\begin{equation}
\delta(\Xi_{i 4}) = \delta(X_{i}^{+ 0} X_{i}^{- +} - X_{i}^{- 0} X_{i}^{+ +}) 
= X_{i}^{+ +}\delta(X_{i}^{- 0} - \frac{X_{i}^{+ 0} X_{i}^{- +}}
{X_{i}^{+ +}})\;,  
\eqnum{3.4c}
\end{equation}

\noindent 
and taking into account the equality
$(1 + \rho_{i})\delta(X_{i}^{- 0} - \frac{X_{i}^{+ 0}X_{i}^{- +}}
{X_{i}^{++}}) = 
\delta[(1 - \rho_{i})(X_{i}^{- 0} - \frac{X_{i}^{+ 0}X_{i}^{- +}}
{X_{i}^{++}})] =
\delta(X_{i}^{- 0} - \frac{X_{i}^{+ 0}X_{i}^{- +}}{X_{i}^{++}})$ 
(coming from the property of the Grassmann variables), the
partition function (2.7) takes the form

\begin{eqnarray}
Z & = &\int {\cal D}X_{i}^{+ +}\;{\cal D}X_{i}^{+ -}\;{\cal D}X_{i}^{- +}\;
{\cal D}X_{i}^{+ 0}\;{\cal D}X_{i}^{0 +}\;
\delta(X_{i}^{+ +} + \frac{X_{i}^{+ -} X_{i}^{- +}}{X_{i}^{+ +}} +
\frac{X_{i}^{0 +} X_{i}^{+ 0}}{X_{i}^{+ +}} - 1) \nonumber \\ 
&\times& exp\;(i \int dt\;L^{*}(X,\dot{X}))\;,
\eqnum{3.5}
\end{eqnarray}

\noindent
where $L^{*}(X,\dot{X})$ is given by

\begin{eqnarray}
L^{*}(X,\dot{X}) & = & \frac{i}{2}\sum_{i} \frac{1}{X^{+ +}_{i}} 
\left(X^{- +}_{i} {\dot X}^{+ -}_{i} - X^{+ -}_{i} {\dot X}^{- +}_{i}\right)
\nonumber \\ 
& + & \frac{i}{2} \sum_{i, \sigma} \frac{1}{X^{+ +}_{i}}\left(X_{i}^{+ 0}
{\dot X}_{i}^{0 +} + X_{i}^{0 +}{\dot X}_{i}^{+ 0}\right) - H(X)\;.
\eqnum{3.6}
\end{eqnarray}

As it is known the change of variables that allows to write the
remaining five Hubbard $X$ variables in terms of the fields
variables in the decoupled slave-fermion representation is defined
by

\begin{equation}
X_{i}^{++} =  b^{\dag}_{i +}b_{i +}\;, 
\eqnum{3.7a}
\end{equation}

\begin{equation}
X_{i}^{+ -} = b^{\dag}_{i +}b_{i -}\;,
\eqnum{3.7b}
\end{equation}

\begin{equation}
X_{i}^{- +} = b^{\dag}_{i -}b_{i +}\;,
\eqnum{3.7c}
\end{equation}

\begin{equation}
X_{i}^{0 +} = b_{i +}f_{i}^{\dag}\;,
\eqnum{3.7d}
\end{equation}

\begin{equation}
X_{i}^{+ 0} = b^{\dag}_{i +}f_{i}\;.
\eqnum{3.7e}
\end{equation}

>From equations (3.7) it can be seen that the five Hubbard $X$-fields are
given in terms of the six fields of the slave-fermion
representation, so it is necessary to introduce an aditional
condition among the six fields of the slave-fermion
representation to make possible the transformation.

We assume the following general linear form for the conditions
in each lattice site 

\begin{equation}
\phi_{i} = \sum_{a}(G_{i a}\,b_{i a} + 
G_{i a}^{\dag}\,b_{i a}^{\dag}) + H_{i}\,f_{i} 
- H^{\dag}_{i}\,f^{\dag}_{i} + K_{i} = 0,
\eqnum{3.8}
\end{equation}

\noindent 
where $G_{i a}$\,,\,$G_{i a}^{\dag}$\,,\,$K_{i}$ are bosonic
parameters and $H_{i}$\,,\, $H_{i}^{\dag}$ are fermionic (Grassmannian)   
parameters. 

As it was commented above, when the Hubbard $X$-operators are
written in a decoupled representation a local gauge symmetry is
made evident. Thus, from a constrained system with a set of
second-class constraints, it changes into a constrained
system with a first-class constraint, and therefore a gauge
fixing condition must be imposed. Therefore, the equation (3.8)
is not other than the gauge fixing condition which corresponds to
the local gauge symmetry appearing in the decoupled representation\cite{15}.

A convenient choice is to take in equation (3.8): $G_{i +} = i$,
$G^{\dag}_{i +} = - i$ and the remaining coeffi\-ci\-ents all zero, i.e  
equation (3.8) reads 

\begin{equation}
\phi_{i} = i(b_{i +} - b_{i +}^{\dag}) = 0\;.
\eqnum{3.9}
\end{equation}

Later on, in equation (3.5) is introduced as usual the unity 

\begin{eqnarray}
1 & = & \int {\cal D}b^{\dag}_{i \sigma}\;{\cal D}b_{i \sigma}\;
{\cal D}f_{i}^{\dag}\;{\cal D}f_{i}\;
\delta(X_{i}^{+ +} - b^{\dag}_{i +}b_{i +})\;
\delta(X_{i}^{+ -} - b^{\dag}_{i +}b_{i -})\;\delta(X_{i}^{- +} 
- b^{\dag}_{i -}b_{i +}) \nonumber \\
&\times& \delta(X_{i}^{0 +} - f_{i}^{\dag} b_{i +})\;
\delta(X_{i}^{+ 0} - b^{\dag}_{i +} f_{i})
\delta(\phi_{i})\;J_{i}\;,
\eqnum{3.10}
\end{eqnarray}

\noindent
where $J_{i}$ is the super Jacobian of the transformation (3.7) and (3.9),
and its value is $J_{i} = (b_{i +} + b_{i +}^{\dag})$.

By integrating out the five variables $X_{i}$, the partition 
function can be written 

\begin{equation}
Z = \int {\cal D}b^{\dag}_{i \sigma}\;{\cal D}b_{i \sigma}\;
{\cal D}f_{i}\;{\cal D}f_{i}^{\dag}\;\delta(\Omega_{i})\;\delta(\phi_{i})\;
J_{i}\;exp\;(i \int dt\; 
L(b_{\sigma}\;,\;b^{\dag}_{\sigma}\;,\;f\;,\;f^{\dag})).
\eqnum{3.11}
\end{equation}

It is easy to see that the super Jacobian $J_{i}$
is equal to minus the determinant of the Dirac bracket
constructed from the first-class constraint $\Omega_{i}$ and the gauge 
fixing condition (3.9), 
i.e $- det[\Omega_{i}\;,\;\phi_{i}]_{D}$, where the first-class 
constraint $\Omega_{i}$ in the slave-fermion representation is given by

\begin{equation}
\Omega_{i} = \sum_{i \sigma} b^{\dag}_{i \sigma}b_{i \sigma} 
+ f_{i}^{\dag} f_{i} - 1 = 0\;.
\eqnum{3.12}
\end{equation}

In the equation (3.11) the Lagrangian 
$L(b_{\sigma}\;,\;b^{\dag}_{\sigma}\;,\;f\;,\;f^{\dag})$ reads

\begin{equation}
L(b_{\sigma}\;,\;b^{\dag}_{\sigma}\;,\;f\;,\;f^{\dag}) =
\frac{i}{2}\sum_{i, \sigma}(b^{\dag}_{i \sigma} \dot{b}_{i \sigma} -
\dot{b^{\dag}}_{i \sigma} b_{i \sigma}) +
\frac{i}{2}\sum_{i} (f_{i}^{\dag}\dot{f_{i}} -
\dot{f_{i}}^{\dag} f_{i}) - H(b^{\dag}_{i \sigma}, b_{i \sigma}, 
f_{i}^{\dag},f_{i})\;.
\eqnum{3.13}
\end{equation}

Therefore, the above considerations show that our correlation generating 
functional (2.7) can be mapped into the correlation
generating functional (3.11) coming from the slave-fermion
representation\cite{15}. 
This mapping is a consequence of the fermionic constraints present in   
our expression for the correlation generating functional (2.7).
We can also conclude that it is not possible to relate our correlation 
generating functional (2.7) with that corresponding to the
slave-boson representation. 

The next question is how to construct from the symplectic FJ formalism 
a new family of first-order Lagrangian by using the 
Hubbard $X$-operators of the graded algebra spl(2,1) as fields 
variables, in such a way that the results can be mapped in the slave-boson 
representation. The problem is solved in the next section

\section{Classical Lagrangian and constraints. Slave-boson representation.}

By following Refs.[18,25], we assume that the family of classical first-order 
Lagrangians in terms of the Hubbard ${\hat X}$-operators can be written as 
follows 

\begin{equation}
L = a_{\alpha \beta}(X){\dot{X}}^{\alpha \beta} - {\bf V^{(0)}}\;.
\eqnum{4.1}
\end{equation} 

\noindent
where the five Hubbard ${\hat X}$-operators $X^{\sigma \sigma'}$
and $X^{0 0}$ are boson-like and the four Hubbard ${\hat X}$-operators 
$X^{\sigma 0}$ and $X^{0 \sigma}$ are fermion-like. In the
present case the symplectic potential is ${\bf V^{(0)}}$ = $H(X)$.

The Lagrangian functional coefficients $a_{\alpha \beta}(X)$
that $a$ $priori$ are unknown must be determined by consistency
in such a way that the graded algebra (2.2) for the
Hubbard ${\hat X}$-operators is verified. 
By following the steps of Ref.[18] it is straightforward to
construct the symplectic supermatrix associated to the Lagrangian (4.1).
Thus the symplectic supermatrix $M_{AB}$ is written in the form

\begin{eqnarray}
M_{AB} =
\left( \matrix {
A_{bb}     &B_{bf}   \cr
C_{fb}     &D_{ff}   \cr }
\right) \; . \eqnum{4.2}
\end{eqnarray}
 
The Bose-Bose parts $A_{bb}$ is a $(10\times10)$-dimensional matrix
and it takes the form

\begin{eqnarray}
A_{bb} =
\left( \matrix {
\frac{\partial a_{\sigma \sigma'}}{\partial X^{\sigma'' \sigma'''}}
- \frac{\partial a_{\sigma'' \sigma'''}}{\partial X^{\sigma \sigma'}}
&\frac{\partial a_{0 0}}{\partial X^{\sigma'' \sigma'''}}
- \frac{\partial a_{\sigma'' \sigma'''}}{\partial X^{0 0}}
&\frac{\partial\Omega_{\sigma \sigma'}}{\partial X^{\sigma'' \sigma'''}}
&\frac{\partial\Omega_{0 0}}{\partial X^{\sigma'' \sigma'''}}    \cr
-\frac{\partial a_{0 0}}{\partial X^{\sigma \sigma'}}
+ \frac{\partial a_{\sigma \sigma'}}{\partial X^{0 0}}
&0
&\frac{\partial\Omega_{\sigma \sigma'}}{\partial X^{0 0}}
&\frac{\partial\Omega_{0 0}}{\partial X^{0 0}}  \cr
- \frac{\partial\Omega_{\sigma'' \sigma'''}}{\partial X^{\sigma \sigma'}}  
&- \frac{\partial\Omega_{\sigma'' \sigma'''}}{\partial X^{0 0}}   
&0 
&0  \cr
- \frac{\partial\Omega_{0 0}}{\partial X^{\sigma \sigma'}}   
&- \frac{\partial\Omega_{0 0}}{\partial X^{0 0}}  &0  &0   \cr }
\right) \; , \eqnum{4.3}
\end{eqnarray}

The Bose-Fermi parts $B_{bf}$ (the Fermi-Bose parts $C_{fb} = - B_{bf}^{T}$)
is a $(4\times10)$-dimensional rectangular supermatrix given by 

\begin{eqnarray}
B_{bf} =
\left( \matrix {
\frac{\partial a_{0 \sigma}}{\partial X^{\sigma'' \sigma'''}}
- \frac{\partial a_{\sigma'' \sigma'''}}{\partial X^{0 \sigma}}         
&\frac{\partial a_{\sigma 0}}{\partial X^{\sigma'' \sigma'''}}
- \frac{\partial a_{\sigma'' \sigma'''}}{\partial X^{\sigma 0}}   \cr
\frac{\partial a_{0 \sigma}}{\partial X^{0 0}}
- \frac{\partial a_{0 0}}{\partial X^{0 \sigma}}         
&\frac{\partial a_{\sigma 0}}{\partial X^{0 0}}
- \frac{\partial a_{0 0}}{\partial X^{\sigma 0}}   \cr
- \frac{\partial\Omega_{\sigma'' \sigma'''}}{\partial X^{0 \sigma}}        
&- \frac{\partial\Omega_{\sigma'' \sigma'''}}{\partial X^{\sigma 0}} \cr
- \frac{\partial\Omega_{0 0}}{\partial X^{0 \sigma}}
&- \frac{\partial\Omega_{0 0}}{\partial X^{\sigma 0}}  \cr }
\right) \; . \eqnum{4.4}
\end{eqnarray}
 
The Fermi-Fermi parts $D_{ff}$ is the $(4\times4)$-dimensional matrix 
given by

\begin{eqnarray}
D_{ff} =
\left( \matrix {
\frac{\partial a_{0 \sigma}}{\partial X^{0 \sigma'}}
+ \frac{\partial a_{0 \sigma'}}{\partial X^{0 \sigma}} 
& \frac{\partial a_{\sigma 0}}{\partial X^{0 \sigma'}}
+ \frac{\partial a_{0 \sigma'}}{\partial X^{\sigma 0}}   \cr
\frac{\partial a_{0 \sigma}}{\partial X^{\sigma' 0}}
+ \frac{\partial a_{\sigma' 0}}{\partial X^{0 \sigma}}
&\frac{\partial a_{\sigma 0}}{\partial X^{\sigma' 0}}
+ \frac{\partial a_{\sigma' 0}}{\partial X^{\sigma 0}}   \cr }
\right) \; , \eqnum{4.5}
\end{eqnarray}

\noindent
where $\Omega_{\sigma \sigma'}$ and $\Omega_{0 0}$
are the appropriate bosonic second-class constraints defining
the structure of the constrained model.

Once the symplectic algorithm is applied and the correspondent
differential equations are solved the solution we found is

\begin{equation}
a_{i 0 \sigma} = \frac{i}{2 X_{i}^{0 0}}\; X_{i}^{ \sigma 0}\;\;,\;\;
\hspace{0.4cm} a_{i \sigma 0} = \frac{i}{2 X_{i}^{0 0}}\; X_{i}^{0 \sigma}\
\eqnum{4.6}
\end{equation}

\noindent
and the boson-like Lagrangian coefficients are all zero.

The set of bosonic second class constraints is given by

\begin{equation}
\Omega_{i}^{0 0} = X_{i}^{0 0} + X_{i}^{+ +} + X_{i}^{- -} - 1 = 0\;,
\eqnum{4.7a}
\end{equation}

\begin{equation}
\Omega_{i}^{\sigma \sigma'} = X_{i}^{\sigma \sigma'} - \frac{X_{i}^{\sigma 0}
X_{i}^{0 \sigma'}}{X_{i}^{0 0}} = 0\;.
\eqnum{4.7b}
\end{equation}

In particular the contraint (4.7a) is the completeness condition
necessary to avoid the double occupancy at each site.

In these conditions the symplectic supermatrix is invertible and
the matrix elements of it inverse gives the correct Hubbard
graded brackets (2.2), i.e

\begin{equation}
(M^{AB})^{-1} = -i (-1)^{\mid\varepsilon_{A}\mid}
\left[{\hat A}\;,\;{\hat B}\right]_{\pm}\;,
\eqnum{4.8}
\end{equation}

\noindent
where ${\mid\varepsilon_{A}\mid}$ is the Fermi grading of the field variable 
$A$.

Consequently the dynamics in this condition is given by the Lagrangian 

\begin{equation}
L(X, \dot{X}) = - \frac{i}{2 X_{i}^{0 0}}
\sum_{i, \sigma}({\dot{X_{i}}}^{0 \sigma}\;X_{i}^{\sigma 0} 
+ {\dot{X_{i}}}^{\sigma 0}\;
X_{i}^{0 \sigma}) - H(X)\;.
\eqnum{4.9}
\end{equation} 

The Lagrangian (4.9) together with the bosonic constraints (4.7)
correspond to a situation in which the bosons are totally
constrained and the dynamics is carried out only by the fermions.
 
The partition function corresponding to this solution reads

\begin{eqnarray}
Z & = & \int {\cal D}X_{i}^{\alpha \beta}\;
\delta[X_{i}^{0 0} + X_{i}^{+ +} + X_{i}^{- -} -1]\;
\delta[X_{i}^{\sigma \sigma'} - \frac{X_{i}^{\sigma 0}
X_{i}^{0 \sigma'}}{X_{i}^{0 0}}]
\;(sdet M_{AB})_{i}^{\frac{1}{2}}\nonumber \\
& \times& exp\;(i \int dt\;L(X, \dot{X}))\;.
\eqnum{4.10}
\end{eqnarray}

By computing the superdeterminant of the symplectic matrix
appearing in (4.10) it results

\begin{equation}
(sdet M_{AB})_{i}^{\frac{1}{2}} = \left(det A \left[det(D - 
CA^{-1}B)\right]^{-1}\right)^{\frac{1}{2}} = (X_{i}^{0 0})^{2}\;.
\eqnum{4.11}
\end{equation}

Now, in order to confront the correlation generating functional (4.10)
with those coming from the slave-boson representation some
algebraic manipulations are needed.  

The first step is to make the following change of variables

\begin{equation}
\varphi_{i 1} = X_{i}^{0 0} - b_{i}^{\dag}\;b_{i} = 0\;,
\eqnum{4.12a}
\end{equation}

\begin{equation}
\varphi_{i \sigma 0} =  X_{i}^{\sigma 0} -  f^{\dag}_{i \sigma}\;b_{i} = 0\;,
\eqnum{4.12b}
\end{equation}

\begin{equation}
\varphi_{i 0 \sigma} =  X_{i}^{0 \sigma} -  f_{i \sigma}\;b_{i}^{\dag} = 0\;.
\eqnum{4.12c}
\end{equation}
 
Analogously to what happens in the slave-fermion representation,
in the decoupled slave-boson one also an additional condition among
the fields is needed. From the general linear equation (3.8) we choose by 
simplicity the following reality condition

\begin{equation}
\varphi_{i 2} = b_{i}^{\dag} - b_{i} = 0\;,
\eqnum{4.12d}
\end{equation}

The super Jacobian $J_{i}$ of the transformation (4.12) is
given by 

\begin{equation}
J_{i} = - \frac{(b_{i}^{\dag} + b_{i})}{(b_{i}^{\dag}b_{i})^{2}}\;.
\eqnum{4.13}
\end{equation}

Now, by introducing in the equation (4.10) the unity

\begin{equation}
1 = \int {\cal D}f^{\dag}_{i \sigma}\;{\cal D}f_{i \sigma}\;
{\cal D}b_{i}^{\dag}\;{\cal D}b_{i}\;J_{i}\;
\delta(X_{i}^{\sigma 0} - f^{\dag}_{i \sigma} b_{i})\;
\delta(X_{i}^{0 \sigma} - f_{i \sigma} b_{i}^{\dag})\;
\delta(b_{i}^{\dag} b_{i} - X_{i}^{0 0})\;\delta(b_{i}^{\dag} - b_{i})\;,
\eqnum{4.14}
\end{equation}

\noindent
and integrating out all the fields variables 
$X_{i}^{\alpha \beta}$, after same algebraic manipulations it is possible 
to show that the partition function (4.10) takes the form

\begin{equation}
Z = \int {\cal D}b_{i}^{\dag}\;{\cal D}b_{i}\;{\cal
D}f^{\dag}_{i \sigma}
{\cal D}f_{i \sigma}\;\;\delta(\Omega_{i})\;\delta(\phi_{i})
\;(b_{i} + b_{i}^{\dag})\;exp\;(i \int dt\; L(b^{\dag}\;,\;b\;,
\;f_{\sigma}^{\dag}\;,\;f_{\sigma}))\;,
\eqnum{4.15}
\end{equation}

\noindent
where $\Omega_{i}$ and $\phi_{i}$ are respectively the first-class
constraint and the gauge fixing condition in the radial gauge\cite{12},
appearing in the partition function of the slave-boson 
representation\cite{17}. They respectively read 

\begin{equation}
\Omega_{i} = b_{i}^{\dag}\;b_{i} + \sum_{i, \sigma} f^{\dag}_{i \sigma}
f_{i \sigma} - 1 = 0\;.
\eqnum{4.16}
\end{equation}

\begin{equation}
\phi_{i} = i(b_{i} - b_{i}^{\dag}) = 0\;.
\eqnum{4.17}
\end{equation}

Again, we note that the factor $(b + b^{\dag})$ in the equation (4.15)
is precisely the value of the $det[\Omega_{i}\;,\;\phi_{i}]_{D}$
appearing in the gauge theories containing first-class constraints. 

The Lagrangian $L(b^{\dag}\;,\;b\;
,\;f_{\sigma}^{\dag}\;,\;f_{\sigma})$ defined in the equation (4.15) 
is given by

\begin{equation}
L(b^{\dag},b, f^{\dag}_{\sigma},f_{\sigma}) =
\frac{i}{2}\sum_{i}(b^{\dag}_{i} \dot{b}_{i} -
\dot{b^{\dag}}_{i} b_{i}) +
\frac{i}{2}\sum_{i, \sigma}(f^{\dag}_{i \sigma}\dot{f}_{i \sigma} -
\dot{f}^{\dag}_{i \sigma} f_{i \sigma}) - H\;. 
\eqnum{4.18}
\end{equation}

In summary, from our approach and working with the Hubbard
$X$-operators without using any decoupling representation,
the new family of Lagrangians (4.9) is obtained. The respective correlation 
generating functional (4.10) is mapped into the solution
provided by the slave-boson representation. It is important to
note that the new path-integral (4.10)
in terms of the Hubbard $X$-operators up to our knowledge was developed  
in the present paper for the first time.

We can see once more how a second-class constrained model
written in terms of the Hubbard $X$-operators, when  
written in terms of the decoupled slave-particle
representations, is transformed 
in a constrained system where a local gauge symmetry is made evident.

We think that the results we can obtain by using the
partition function (4.10) with the Lagrangian (4.9) can be
useful for regimes where the system is close to a Fermi liquid state.
 
In a forthcoming paper the partition function (4.10) is studied in detail  
within the context of the perturbative formalism.
Having in mind the difficulty to treat the first-class constraint inside the 
path-integral (4.15), our purpose is to construct
the Feynman rules and the diagrammatics starting from the
path-integral (4.10). Once an appropriate fermion 
propagator can be found, our first objective will be to analyse 
the properties of the fermion spectral function.

\section{Two alternative ways to define the propagator of fermion modes.}
 
In this section we discuss two alternative ways to treat the
fermionic sector in order to define the propagation of the fermion
modes. As commented above, a crucial problem in the t-J
model is to define the fermionic propagation in the constrained
Hilbert space, when the double occupancy is forbidden.

With the purpose to study this problem, in Ref.[19] the
correlation gene\-ra\-ting functional (2.12) 
was considered at finite temperature
by means of the "Euclideanization procedure".
Moreover it was assumed that we are close to an undoped regime
where the system is an antiferromagnetic insulator. 
Under this condition there is a small
number of holes and it can be assumed that
the hole density $\rho_{i} = <\rho_{i}> = constant$.
The constant value $\rho$ of the hole density must
be determined later on by consistency, for a given value of the chemical
potential $\mu$. 

In these conditions, it is possible to treat the non-polynomic Lagrangian 
(2.13) in the framework of the
perturbative formalism, and so it can be partitioned as follows 
 
\begin{equation}
L_{eff} = L^{B}({\bf {S}}, \lambda) + L^{F}({\mbox{\boldmath{$\eta$}}}) +
L^{I}({\bf {S}},{\mbox{\boldmath{$\eta$}}})\;,
\eqnum{5.1}
\end{equation}

\noindent
where

\begin{eqnarray}
L^{B}({\bf {S}}, \lambda) & = & - \frac{i}{2s} 
\sum_{i} \frac{{\widetilde S}_{i1}{\dot{\widetilde S}_{i2}} 
- {\widetilde S}_{i2}{\dot{\widetilde S}_{i1}}}{s + s'} +  
2 s' \sum_{i} \lambda_{i}{\widetilde S}_{i3} \nonumber \\
& + & \frac{1}{8 s^{2}}\sum_{i,I} J' \left[
{\widetilde S}_{i1}{\widetilde S}_{(i +I)1}
- {\widetilde S}_{i2}{\widetilde S}_{(i +I)2} 
- {\widetilde S}_{i3}{\widetilde S}_{(i +I)3} 
+ {\widetilde S}_{i}^{2}\right] \;,
\eqnum{5.2a}
\end{eqnarray}
 
\noindent
and

\begin{eqnarray}
L^{F}({\mbox{\boldmath{$\eta$}}}) + L^{I}({\bf S}, 
{\mbox{\boldmath{$\eta$}}}) & = & \sum_{i, \sigma} 
{\dot{\Psi}^{*}_{i \sigma}} {\Psi_{i \sigma}} +  
\mu \sum_{i, \sigma}{\Psi_{i \sigma}}{\Psi^{*}_{i \sigma}} 
+ \sum_{i,j, \sigma} t_{ij} {\Psi}_{i \sigma} 
{\Psi}^{*}_{j {\bar{\sigma}}} \nonumber \\
& + & \sum_{i}
{\bar {\mbox{\boldmath{$\eta$}}}}_{i}\;{\cal M}_{i}\;
{\mbox{\boldmath{$\eta$}}}_{i}\;.  
\eqnum{5.2b}
\end{eqnarray}

Considering the bilinear bosonic part of the equation (2.13), for a 
constant value of the hole density and taking $J_{ij} = constant$, 
we arrive at the equation (5.2a). In the equation (5.2a) was defined 
$J' = J(1 - \rho)^{2}$. Moreover, the symbol $\sum_{I}$ indicates sum over 
nearest-neighbor sites.

Besides, in equation. (5.2a) it was assumed that the vector ${\bf S}$ 
is written

\begin{equation}
{\bf S} = (0,0,s') + ({\widetilde {S_{1}}}\;,\; {\widetilde {S_{2}}}\;,
\;{\widetilde {S_{3}}})    
\eqnum{5.3}
\end{equation}

\noindent 
where ${\widetilde {S_{1}}}\;,\; {\widetilde {S_{2}}}\;,
\;{\widetilde {S_{3}}}$ are the fluctuations. 
So, the equation (5.2a) corresponds to the lowest order of the
system fluctuating around on antiferromagnetic state.
Moreover, we must consider $s^{'} \neq s$ because as known the local 
magnetisation in an antiferromagnetic state reduced from its classical value, 
even for the pure Heisenberg model. The value of $s^{'}$ must be
determined also by consistency.

In equation (5.2b) the four-component spinor
${\mbox{\boldmath{$\eta$}}} 
= \left( \matrix {
{\mbox{\boldmath{$\Psi$}}}   \cr
{\mbox{\boldmath{$\xi$}}}   \cr }\right)$, 
is constructed from the two 
spinors ${\mbox{\boldmath{$\Psi$}}}$
and ${\mbox{\boldmath{$\xi$}}}$. 
The physical two-component spinor ${\mbox{\boldmath{$\Psi$}}} = 
\left( \matrix {
\Psi_{+}   \cr
\Psi_{-}   \cr }\right)$ is restricted by the fermionic constraint equations
(2.10c,d), and the two-component spinor ${\mbox{\boldmath{$\xi$}}} = 
\left( \matrix {
\xi_{+}   \cr
\xi_{-}   \cr }\right)$ is a Majorana spinor.

In the same equation the $4\times4$ dimensional matrix ${\cal M}$ is defined 
by 

\begin{eqnarray}
{\cal M} =
\left( \matrix {
0     &{\bf I}s + {\bf{S}}.{\mbox{\boldmath{$\sigma$}}}   \cr
{\bf I}s + {\bf{S}}.{\mbox{\boldmath{$\sigma$}}}       &0   \cr }
\right) \; , \eqnum{5.4}
\end{eqnarray}

\noindent
where ${\mbox{\boldmath{$\sigma$}}} = (\sigma_{1},\sigma_{2}, \sigma_{3})$ 
are the Pauli matrices.

In this regime the diagrammatics and the Feynman rules can be found. 
In parti\-cu\-lar the bilinear part of the bosonic sector
written in equation (5.2a) gives rise 
to the usual antiferromagnetic magnon propagator (see Ref.[19]).

The bilinear fermionic part $L^{F}({\mbox{\boldmath{$\eta$}}})$ of 
the equation (5.2b) can be written in terms of the four-component spinor 
${\mbox{\boldmath{$\eta$}}}$ and it is given by
 
\begin{equation}
L^{F}({\mbox{\boldmath{$\eta$}}}) = \sum_{i,j}{\bar{\eta}_{i \alpha}}\; 
(G^{-1}_{(0) ij})^{\alpha \beta}\;{\eta}_{j \beta} \;,
\eqnum{5.5}
\end{equation}

\noindent
where in the Fourier space the symmetric non-singular $4\times4$ 
dimensional matrix $G^{-1}_{(0)}$ is defined by

\begin{eqnarray}
(G^{\alpha \beta}_{(0)})^{-1}(k, \nu_{n}, \nu_{n}^{'}) = 
\left( \matrix {
- (i \nu_{n} + \mu)   &\varepsilon_{k}
&\frac{1}{2}(s + s')             &0   \cr
\varepsilon_{k}  &- (i \nu_{n} + \mu)
&0              &\frac{1}{2}(s - s')   \cr
\frac{1}{2}(s + s')          &0         &f        &g            \cr
0     &\frac{1}{2}(s - s')              &g        &- f            \cr }
\right) \; \delta(\nu_{n}, \nu_{n}^{'})\;. \nonumber \\ 
\eqnum{5.6}
\end{eqnarray}
  
In Eq.(5.6) the quantities $k$ and $\nu_{n}$ are respectively the momentum 
and the Matsubara frequency of the fermionic field, and was  
defined $\varepsilon_{k}= - t \,\sum_{I} exp(-i{\bf I.k})$.

The functions $f$ and $g$ appearing in Eq. (5.6) are totally arbitrary.
As easily can be seen, these functions do not appear in
the Lagrangian due to the Majorana condition on the two-component
spinor ${\mbox{\boldmath{$\xi$}}}$. 

In this scenary, the symmetric matrix defining the fermionic free propagator 
$G_{(0) \alpha \beta}$ is given by the inverse of the matrix (5.6). The
physical components of the free propagator are given by the matrix
elements $G_{(0) 11}$,$G_{(0) 12}$,$G_{(0) 21}$ and $G_{(0) 22}$
and they were explicitly written and analysed in section IV. of Ref.[19].
So, the Feynman rules propagators and vertices are given straightforward
and therefore the boson and fermion self-energy can be computed.
 
However, some particular features of the fermionic free propagator must be 
emphasised. 

The trick of introducing an auxiliary two-component Majorana spinor is a 
way to obtain a free functional $G_{(0)}$ that really propagates physical 
fermionic modes. 
 
The electron spectral function is defined from the fermionic
propagator $G_{(0) \alpha\beta}$ by considering the components 
$G_{(0) 11}$ and $G_{(0) 22}$.
The matrix elements directly connected 
with the electronic properties, as for example the Fermi surfase (FS), are 
precisely $G_{(0) 11}$ and $G_{(0) 22}$. The electronic spectral function
measured in photoemission experiments\cite{26} must be related with minus
the imaginary part of these matrix elements.

Contrary to the fermion propagator obtained by means of the standard 
Green function method\cite{27}, our fermion propagator
contains two poles. It is important to say that for a given filled factor
the chemical potential we obtain is exactly the same obtained 
by using the standard Green function method.

By ploting the electron spectral function (see Fig. 1 Ref.[19]) 
we can see that the peak at negative energy must be interpreted   
as the extraction of an electron, while the peak at positive energy 
represents the addition of an electron to the system. Therefore, the two peaks
account for the photoemission and the inverse of the photoemission 
respectively. The presence of these two peaks implies that for a given value
of $k$, the state is not completely filled or empty. Photoemission
experiments are only sensitive to the first peak. Then the first peak of our 
propagator must be related with the excitation measures in photoemission.

An alternative way to treat the fermionic sector is to
start from the equation (2.7) and to work inside the path-integral.
So, by inte\-gra\-ting out the two delta functions on the fermionic 
constraints $\Xi_{i3}$ and $\Xi_{i4}$, the partition function can be written 
as follows

\begin{equation}
Z = \int {\cal D}X_{i}^{\sigma \sigma'}\;{\cal D}X_{i}^{0 +}
{\cal D}X_{i}^{+ 0}\;\delta(\Omega_{i1})\;\delta(\Omega_{i2})\;  
(sdet M_{AB})_{i}^{\frac{1}{2}}\;exp\;i \int dt\; L^{*}(X,\dot{X}) \;,
\eqnum{5.7}
\end{equation}

\noindent
where $L^{*}(X,\dot{X})$ is given by

\begin{eqnarray}
L^{*}(X,\dot{X}) & = & \frac{i}{2}\sum_{i} \frac{1}{X^{+ +}_{i}} 
\left(X^{- +}_{i} {\dot X}^{+ -}_{i} - X^{+ -}_{i} {\dot X}^{- +}_{i}\right)
\nonumber \\ 
& + & \frac{i}{2} \sum_{i, \sigma} \frac{1}{X^{+ +}_{i}}\left(
{\dot X}_{i}^{0 +} X_{i}^{+ 0}
+ {\dot X}_{i}^{+ 0} X_{i}^{0 +}\right) - H(X)\;.
\eqnum{5.8}
\end{eqnarray}

The total Hamiltonian $H$ is defined by

\begin{equation}
H = H_{t-J} + \mu \sum_{i, \sigma}\;X_{i}^{0 \sigma}\;X_{i}^{\sigma 0}\;,     
\eqnum{5.9}
\end{equation}

\noindent
where the Hamiltonian $H_{t-J}$ defined in equation (2.9) must be evaluated 
on the fermionic constraints $\Xi_{i3}$ and $\Xi_{i4}$. 

Due to the non-linearity of the constraints (2.10c,d), when the
path-integration on the two fermionic fields $X_{i}^{0 -}$ and
$X_{i}^{- 0}$ is carried out, the non-polynomial structure of
the kinetic fermionic part of the Lagrangian is made evident,
as it can be seen from the equation (5.8).

After the four boson-like $X$-Hubbard operators are related to
the real components $S_{\alpha}$ ($\alpha = 1,2,3$) of a vector field 
${\bf S}$  and the remaining two fermion-like 
$X$-Hubbard operators written
in terms of suitable component spinor fields (see equations (2.11)), 
the correlation generating functional (5.7) takes the form

\begin{equation}
Z  =  \int  {\cal D} S_{i1} \; {\cal D} S_{i2}\; {\cal D} S_{i3}\; 
{\cal D}{\Psi}_{i -}\; 
{\cal D}{\Psi}^{*}_{i -}\; 
{\cal D}{\lambda_{i}}\;(sdet M_{AB})_{i}^{\frac{1}{2}}\;
\left(\frac{\partial X}{\partial S}\right)_{i}\;
exp\;(i \int\; dt \; L_{eff})\;.
\eqnum{5.10}
\end{equation}

Now, the Lagrangian $L_{eff}$ defined in equation (5.10) is given by

\begin{eqnarray}
L_{eff} & = & \frac{1}{2 s} \sum_{i} (1 - \rho_{i})  
\left(\frac{ S_{i 1} {\dot S}_{i 2} - S_{i 2} {\dot S}_{i 1}}{s
+ S_{i 3}}\right) + is \sum_{i}\frac{1}{s + S_{i 3}}
\left({\dot{\Psi}}_{i -}^{*} \Psi_{i -}
+ {\dot {\Psi}}_{i -}\Psi_{i -}^{*}\right)\nonumber \\
& - & H + \sum_{i} \lambda_{i}(S_{i 1}^{2} +S_{i 2}^{2} 
+S_{i 3}^{2} - s^{2})\;.
\eqnum{5.11}
\end{eqnarray}

\noindent
where in equation (5.11) and hereafter the tilde over the fluctuations 
is omitted.

The Hamiltonian $H$ written in terms of the real vector 
variable ${\bf S}$ and the two spinor component fields $\Psi_{-
i}$ and $\Psi_{- i}^{*}$, reads

\begin{eqnarray}
H & = & \sum_{i,j}\frac{t_{ij}}{(s + S_{i 3})(s + S_{j 3})}
\left[S_{i 1}S_{j 1} +S_{i 2}S_{j 2} +S_{i 3}S_{j 3} + s^{2} +
s(S_{i 3} + S_{j 3})\right.\nonumber \\ 
& + & \left.i(S_{i 1}S_{j 2} -S_{i 2}S_{j 1})\right] \Psi_{- i}\Psi_{- j}^{*}
+\frac{1}{8 s^{2}}\sum_{i,j} J_{ij}(1 - \rho_{i})(1 - \rho_{j}) \left[
S_{i 1}S_{j 1} + S_{i 2}S_{j 2}\right.\nonumber \\ 
& + & \left. S_{i 3}S_{j 3} - s^{2}\right]
 + 2s{\mu} \sum_{i}\left(\frac{1}{s + S_{i 3}}\right)
\Psi_{- i}\Psi_{- i}^{*} \;.  
\eqnum{5.12}
\end{eqnarray}
 
Again, the path-integral (5.8) is considered in the framework
of the perturbative formalism at finite temperature, and we
assume that we are close to an undoped regime (antiferromagnetic
insulator).

After a rotation of spins on the second sublattice by $180^{0}$
about the $S_{1}$ axis is performed 
the Euclidean and rotated Lagrangian $L^{ER}_{eff}$ is obtained, and so  
the lowest order of the effective Lagrangian (5.11)
can be partitioned as follows
 
\begin{equation}
L_{eff}^{ER} = L^{B}({\bf {S}}, \lambda) + L^{F}
(\Psi_{- i},\Psi^{*}_{- i}) +
L^{I}({\bf {S}},\Psi_{- i},\Psi^{*}_{- i})\;,
\eqnum{5.13}
\end{equation}

\noindent
where

\begin{eqnarray}
L^{B}({\bf {S}}, \lambda) & = & - \frac{i}{2s}(1 - \rho) 
\sum_{i}\frac{{S}_{i1}{\dot{S}_{i2}} 
- {S}_{i2}{\dot{S}_{i1}}}{s + s'} +  
2 s' \sum_{i}\lambda_{i}{S}_{i3} \nonumber \\
& + & \frac{1}{8 s^{2}}\sum_{i,I} J' \left[
{S}_{i1}{S}_{(i +I)1}
- {S}_{i2}{S}_{(i +I)2} 
- {S}_{i3}{S}_{(i +I)3} 
+ {S}_{i}^{2}\right] \;,
\eqnum{5.14a}
\end{eqnarray}
 
\noindent
and

\begin{equation}
L^{F} = \frac{s}{s + s'} \sum_{i}({\dot {\Psi}}_{i -}^{*}\Psi_{i -}
 + {\dot {\Psi}}_{i -}\Psi_{- i}^{*}) +
\frac{2 \mu s}{s + s'}\sum_{i} \Psi_{i -} \Psi^{*}_{i -}
\eqnum{5.14b}
\end{equation}

\begin{equation}
L^{I} = \sum_{i,j}\frac{t_{ij}}{s +s'}\left({S}_{i1}
- i {S}_{i2} + {S}_{j1} + i {S}_{j2}\right)
\Psi_{i -} \Psi^{*}_{j -} 
\eqnum{5.14c}
\end{equation}
 
At this stage it is important to note that our effective theory
does not contain fermion dispersion. This feature is also
present in the slave-fermion theories when the fermion dynamics
is generated $via$ de interaction with virtual magnons.

By making a Fourier transformation it is possible
to see that the bilinear bosonic part of the Lagrangian
(5.14a) allows to recover the structure of the bosonic
propagator (antiferromagnetic magnons) given by

\begin{eqnarray}
{\cal D}^{ab}_{(0)}(q, \omega_{n}, \omega_{n}^{'}) = 
\left( \matrix {
\frac{J'z}{4 s^{2} d_{(0)}}(1 - \gamma_{q})  & - \frac{2 \omega_{n}}
{(s + s') d_{(0)}}(1 - \rho)              &0              &0   \cr
\frac{2 \omega_{n}}{(s + s') d_{(0)}}(1 - \rho)  &\frac{J'z}{4 s^{2} d_{(0)}}
(1 + \gamma_{q})              &0              &0   \cr
0         &0         &0        &\frac{1}{2s'}            \cr
0     &0   &\frac{1}{2s'}    & - \frac{J'z(1 - \gamma_{q})}{16 s^{2}s'^{2}}
                           \cr }
\right) \; \delta(\omega_{n}, \omega_{n}^{'})\;. \nonumber \\ 
\eqnum{5.15}
\end{eqnarray}

In Eq.(5.15) the quantity $d_{(0)}$ is given by

\begin{equation}
d_{(0)} = \left(\frac{2(1 - \rho)}{s + s'}\right)^{2}
\left(\omega_{q}^{2} + \omega_{n}^{2} 
\right) \;, 
\eqnum{5.16}
\end{equation}

\noindent
where the frequency $\omega_{q}^{2}$ is defined by 

\begin{equation}
{\omega_{q}}^{2} = \left[\frac{z J'}{4 s^{2}} \left(\frac{s + s'}{2(1 - \rho)}
\right)\right]^{2}\;(1 - \gamma_{q}^{2})\;.
\eqnum{5.17}
\end{equation}

Moreover, in Eq.(5.15) $z$ is the number of nearest neighbors
and was defined the quantity $\gamma_{q} = \frac{1}{z}\sum_{I}
exp(i{\bf I.q})$.

Analogously, the bilinear fermionic part (5.14b) reads

\begin{equation}
L^{F} = \sum_{k, \nu_{n}} \Psi^{*}_{-}(k, \nu_{n})\;
G_{0}^{-1}\; \Psi_{-}(k, \nu_{n})\;,
\eqnum{5.18}
\end{equation}

\noindent
where we have named

\begin{equation}
G_{0}^{-1} = \frac{2 s}{s + s'}(i \nu_{n} - \mu)\;.
\eqnum{5.19}
\end{equation}

The inverse of this scalar function given by 

\begin{equation}
G_{0} = \frac{s + s'}{2 s} \frac{1}{i \nu_{n} - \mu}\;,
\eqnum{5.20}
\end{equation}

is a (non-propagating) functional which only depends on the Matsubara 
frequency $\nu_{n}$.

Finally, in the approximation that we consider the unique three-leg vertex is
defined by

\begin{eqnarray}
U_{a} =  \frac{1}{s + s'} 
\left( \matrix {
\varepsilon(k') + \varepsilon(k)   \cr
i(\varepsilon(k') - \varepsilon(k))   \cr
- \frac{s}{s + s'} [i(\nu + \nu') - 2 \mu]          \cr
0          \cr }
\right) \; \delta(q + k - k')\;\delta(\omega + \nu - \nu')\;. \nonumber \\ 
\eqnum{5.21}
\end{eqnarray}

At this point, the problem is to analyse the bilinear 
fermionic sector, in order to give the prescriptions for the propagation 
of the fermionic modes. The usual way to solve the propagation of fermions 
is by means of the Dyson equation. As known the Dyson theorem allows to compute
the inverse of the corrected fermion propagator in terms of the
free fermion propagator and the self-energy. Therefore the propagator  
$G(k, \nu_{n}) = [G_{0}^{-1}(\nu_{n}) - \Sigma(k, \nu_{n})]^{-1}$ 
can be calculated in a straightforward way within the self-consistent
Born approximation\cite{28,29}. By using standard techniques the
following expression for the self-energy at zero temperature is found

\begin{eqnarray}
\Sigma(k, i\nu_{n}) & = & \frac{(1 + \rho)}{2 N}\;t^{2}\;z^{2}
\sum_{q}\frac{\left[\gamma_{k}[1 - (1 - \gamma_{q}^{2})^{1/2}]^{1/2} - 
\gamma_{k + q}[1 + (1 - \gamma_{q}^{2})^{1/2}]^{1/2} \right]^{2}}{
(1 - \gamma_{q}^{2})^{1/2}}
\nonumber \\
&\times& \frac{1}{i \nu_{n} - \omega_{q} - \mu - \Sigma(k +
q\;,\;i \nu_{n} - \omega_{q})}\;
\eqnum{5.22}
\end{eqnarray}

The expression (5.22) is useful in the strong coupling case $(t>J)$. 
On the other hand the self-consistent solution of this
equation is necessary in order to obtain fermionic propagation,
and it must be performed numerically. Once an appropriate
self-energy function $\Sigma(k, i\nu_{n})$ is found the propagator 
$G(k,\nu)$ remains
well defined and it is possible to compute numerically the
spectral function defined by $A(k,\nu) = - \frac{1}{\pi}
lim_{\varepsilon \rightarrow 0} G(k,\nu + i\varepsilon)$.
Finally, as well known, the correction to the bosonic propagator
is given by

\begin{equation}
{\cal D}_{ab} = \left[{\cal D}_{(0) ab}^{-1} - \Pi_{ab}\right]^{-1}\;,
\eqnum{5.23}
\end{equation}

\noindent
where ${\cal D}_{(0) ab}^{-1}$ is the inverse of the 
free bosonic propagator (5.15) and the bosonic self-energy $\Pi_{ab}$
reads

\begin{equation}
{\Pi}_{ab}(q\;,\omega_{n}) = \left(\frac{s + s'}{2 s}\right)^{2}
\sum_{k,\nu_{n}} \frac{U_{a}\;U_{b}}{[i \nu_{n} 
+ \mu - \Sigma(k,\nu_{n})]
[i(\nu_{n} + \omega_{n}) + \mu - \Sigma(q + k\;,\;\nu_{n} + \omega_{n})]}\;.
\eqnum{5.24}
\end{equation}

As it can be seen from the above equation the dimension of the underlying 
lattice and the physics depend on the parameters $z$, $\gamma_{q}$ and 
$\varepsilon_{k}$, though our general formalism is dimensional 
independent.

It is important to confront our results with
other previously given in the literature related with the
spin-polaron theories\cite{28}. Like in these theories our
starting point was to assume an antiferromagnetic order state. This
physical assumption is directly connected with the fact that at
lowest order the fermion is not propagating (see equation
(5.20)). Then, in order to describe a metallic phase where the
holes move coherently on the lattice, it is necessary to solve 
the self-consistent equation (5.22).

The solution of our equations (5.22) and (5.24) together with a
quantitative comparison with the spin-polaron theories is an
important matter that deserves further study. 

Another point to take into account in a future work is to study
the relationship between our matricial propagator $G_{(0) \alpha \beta}$
(see equation (4.13) of Ref[19]) and those obtained by solving 
self-consistently the equation (5.22).

As it is known, any model or approach will be considered as a
good candidate to describe high $T_{c}$ superconductors when it is 
able to answer the question of why the 
antiferromagnetic long range order disappears for small values of
doping (for intance $\rho = 0.04-0.05$). In the last years
this problem was attacked from differents approaches\cite{13,14,30}.
In a future work and from our formalism, we will also have a
response to give about this important point related with the disappearance 
of the antiferromagnetism.

In the present section, the magnetic excitations that we have
considered are antiferromagnetic magnons and in addition  we have assumed a
strong long range antiferromagnetic order. Therefore,
our next step must be to study the instability of the
antiferromagnetic order and to analyse against which phase this is unstable.
In order to have some idea about this fact, in Ref.[19] we
have studied the magnon self energy effects on the magnetic
spectral function. Besides the sofltening of the
antiferromagnetic magnon we have also found a reduction of the
magnetic spectral signal. These results were got by using our
two pole bare fermionic propagator. In the near future and in order 
to improve our calculation we will solve a self
energy coupled problem for both magnetic and electronic dynamics.

\section{Conclusions}

As showed first in Ref.[25] for the pure bosonic case (su(2)
algebra), in a classical Lagrangian formalism it is not possible to 
introduce the full Hubbard algebra by means of constraints.
Consequently, in a path-integral formulation it cannot be
introduced the complete information of the Hubbard algebra namely,
the commutation rules, the completeness condition and the
multplication rules for the Hubbard $X$-operators. So, the
Heisenberg model treated in the Lagrangian picture only admits
two second-class constraints, and these are the completeness
condition $X^{+ +} + X^{- -} + X^{0 0} = 1$ and the non-linear 
constraint $X^{+ -}\,X^{- +} + \frac{1}{4}
(X^{+ +} - X^{- -})^{2} = s^{2}$.

The above last constraint in not really the group Casimir operator.
It can be shown that the presence of such constraint is
consistent with the quantisation of a spin system in the limit
of large spin s, or equivalently for magnetic order state.

A similar situation actually occurs in the case in which the
Hubbard $X$-operators verify the graded algebra spl(2,1), but in
this case at least two solutions are possible.
When the Hubbard $X$-operators closes the graded algebra spl(2,1)
the t-J model described in terms of a first-order Lagrangian has
the following possible solutions:

i) One is the family of first-order Lagrangians (2.8) together with the
set of second-class constraints (2.10). Two of them are bosonics
and the other two are fermionics. In particular the constraint
(2.10a) is the completeness condition. As it was shown the
corresponding path-integral formalism is mapped into the
decoupled slave-fermion representation. So, in this case our
correlation generating functional (2.7) privilege the magnon
dynamics of the system with a strong feature of magnetic
order state (consistent with the large s nature of the
constraint (2.10b)).

ii) The different family of first-order Lagrangians (4.9)
together with the new set of second-class constraints (4.7) is
also a possible solution. In this case all the constraints are bosonic, 
and (4.7a) is again the completeness condition. As it can be
seen the remaining four constraints (4.7b) are related to the
multiplication rules. In this situation the bosons are totally
constrained and the dynamics is carried out only by the fermions. 
Moreover, we note that not any non-linear constraint of the type (2.10b) 
appears. 
As it was shown the path-integral formalism corresponding to this
dynamical situation is mapped in the slave-boson representation.
Therefore, in our correlation generating functional (4.10) the
fermion dynamics with a strong feature of Fermi liquid is priveleged.

It is possible to conclude that once the set of second class-constraints
is chosen, different families of Lagrangians are obtained, and
so we can insure that each family contains different physics.

Worthwhile to remark that both Lagrangian formalism are 
independent of the dimension of the underlying lattice.

Moreover, as it was seen in all the cases the completeness condition appears 
as necessary. As it is well known the completeness condition involve an 
important physical meaning. Such condition avoids at the quantum level the 
double occupancy at each lattice site.

Finally, in section V two alternative way to define the fermion
propagator were developed. By means of the trick of introducing 
auxiliary Majorana spinors, a free fermion matricial propagator
having two poles was found. Later on, by integrating out two of
the fermions using the delta functions, it was possible to
obtain  the non-propagating scalar function (5.20) in the fermionic sector. 
>From this free "propagator" and by means of the Dyson equation
the fermion self-energy can be evaluated straightforward within
the self-consistent Born approximation.

Since our path-integral formalism is mapped in the
slave-fermion formalism and taking into account that the equation
(5.22) for the self-energy at zero temperature has a similar
structure to that obtained from the spin-polaron theories\cite{28},
both results allows us to insure that our approach is consistent.

\end{document}